\documentstyle[12pt]{article}
\newcommand{\beq}{\begin{equation}}
\newcommand{\eeq}{\end{equation}}
\newcommand{\bea}{\begin{eqnarray}}
\newcommand{\eea}{\end{eqnarray}}

\newcommand{\k}{\kappa}

\begin{document}
\setcounter{page}{0}
\topmargin 0pt
\oddsidemargin 5mm
\renewcommand{\thefootnote}{\fnsymbol{footnote}}
\newpage
\setcounter{page}{0}
\begin{titlepage}
\begin{flushright}
QMW 94-37
\end{flushright}
\begin{flushright}
hep-th/9410nnn
\end{flushright}
\vspace{0.5cm}
\begin{center}
{\large {\bf Relevant Perturbations Of The $SU(1,1)/U(1)$ Coset}} \\
\vspace{1.8cm}
\vspace{0.5cm}
{\large Oleg A. Soloviev
\footnote{e-mail: soloviev@V1.PH.QMW.ac.uk}\footnote{Work supported by the
PPARC and in part by a contract from the European Commission Human Capital
and Mobility Programme.}}\\
\vspace{0.5cm}
{\em Physics Department, Queen Mary and Westfield College, \\
Mile End Road, London E1 4NS, United Kingdom}\\
\vspace{0.5cm}
\renewcommand{\thefootnote}{\arabic{footnote}}
\setcounter{footnote}{0}
\begin{abstract}
{It is shown that the space of cohomology classes of the $SU(1,1)/U(1)$ coset
at negative level $k$ contains states of relevant conformal dimensions. These
states correspond to the energy density operator of the associated nonlinear
sigma model. We exhibit that there exists a subclass of relevant operators
forming a closed fusion algebra. We make use of these operators to perform
renormalizable
perturbations of the $SU(1,1)/U(1)$ coset. In the infra-red limit, the
perturbed theory flows to another conformal model.
We identify one of the
perturbative conformal points with the $SU(2)/U(1)$ coset at positive level.
{}From the point of view of the string target space geometry, the given
renormalization group flow maps the euclidean black hole geometry described by
the $SU(1,1)/U(1)$ coset into the sphere described by the $SU(2)/U(1)$ coset.}
\end{abstract}
\vspace{0.5cm}
\centerline{October 1994}
 \end{center}
\end{titlepage}
\newpage
\section{Introduction}

It has been realized that nonunitary Wess-Zumino-Novikov-Witten (WZNW) models
play a significant role in string theory \cite{Lykken}-\cite{Dijkgraaf}. In
some
sense these models appear to be more fundamental than ordinary unitary WZNW
models. We have recently exhibited that the latter can be obtained from
nonunitary WZNW models through renormalization group flows
\cite{Soloviev1},\cite{Soloviev2}. The curious fact about these flows is that
in the presence of two dimensional quantum gravity cosmological minkowskian
string solutions can smoothly flow to euclidean string solutions, whenever the
corresponding CFT's admit proper space-time interpretations \cite{Soloviev3}.

However, nonunitary WZNW models, which are based on nonunitary affine groups,
turn out to be very complicated systems because their spectra contain states of
negative norm. Therefore, in order for these models to make sense, one has to
find a certain way to get rid of negative normed states. Witten has proposed
that WZNW models on compact groups at negative level can be understood via
analytic continuation to noncompact groups \cite{Witten1},\cite{Witten2}. It
has been argued that nonunitary states of the latter can be eliminated by a
coset construction $G/H$, where $G$ is a noncompact group and $H$ is its
maximal compact subgroup \cite{Dixon}.

As yet the problem of nonunitary WZNW models has not been solved completely.
Perhaps, the $SU(1,1)/U(1)$ coset is the only case which has been studied in
detail \cite{Dixon}. It turns out that unitary $N=2$ superconformal models as
well as physical states of two dimensional black holes can be extracted from
unitary representations of $SU(1,1)$
\cite{Lykken},\cite{Dixon},\cite{Witten1},\cite{Dijkgraaf}. These unitary
representations are described as the $SU(1,1)/U(1)$ coset with $U(1)$ being the
compact subgroup of $SU(1,1)$.

The aim of the present paper is to explore relevant perturbations on the
$SU(1,1)/U(1)$ gauged WZNW model. We will exhibit that the physical subspace of
cohomological classes of the $SU(1,1)/U(1)$ gauged WZNW model has relevant
conformal operators corresponding to highest weight vectors of unitary Virasoro
representations. At the same time, these unitary states are descendants of
highest weight states of nonunitary (finite dimensional) representations
of the affine group. These unitary Virasoro representations have been missed in
the previous analysis of the $SU(1,1)/U(1)$ coset \cite{Dixon},\cite{Bars}.
A nongauged version of these representations has been discussed in
\cite{Soloviev4}. We will show that these new relevant operators obey a fusion
algebra which allows us to use them to perform renormalizable perturbations on
the $SU(1,1)/U(1)$ coset. These perturbations are different from canonical
perturbations
of gauged WZNW models considered in \cite{Soloviev2} (see also \cite{Ahn}).
The new type of relevant
perturbations to be discussed in this paper provides renormalization group
flows between noncompact target space geometries (like euclidean black holes)
and compact geometries. This is, in fact, a new sort of topology change
generated by relevant quasimarginal
conformal operators but not (truly) marginal as in the
case of Calabi-Yau manifolds \cite{Aspinwall}.

The paper is organized as follows. In section 2 we will construct new relevant
conformal operators which belong to $\mbox{Ker}Q/\mbox{Im}Q$ of the
$SU(1,1)/U(1)$ gauged WZNW model at negative level $k$. Here $Q$ is the
corresponding BRST operator. In section 3 we will study the fusion rules of
these BRST invariant relevant operators. In particular we will exhibit a
subclass of operators which form closed fusion algebras. In section 4 we will
apply the relevant
operators to perform renormalizable perturbations of the $SU(1,1)/U(1)$ coset
in the limit $k\to-\infty$. We will discuss the renormalization group flow from
the $SU(1,1)/U(1)$ coset to the infra-red conformal point. Finally, in the last
section we will summarize our results and comment on them.

\section{$SU(1,1)/U(1)$ coset}

Let us consider the level $k$ WZNW model defined on the group manifold $G$
corresponding to the Lie group $G$. The action of the theory is given as
follows \cite{Witten3},\cite{Knizhnik}.
\begin{equation}
S_{WZNW}(g,k)=-{k\over4\pi}~\int\left[\mbox{Tr}|g^{-1}\mbox{d}g|^2~+~
{i\over3}\mbox{d}^{-1}\mbox{Tr}(g^{-1}\mbox{d}g)^3\right],\end{equation}
where $g$ is the matrix field taking its values on the Lie group $G$. For
compact groups the Wess-Zumino term \cite{Wess} is well defined only modulo
$2\pi$ \cite{Witten3}, therefore, the parameter $k$ must be an integer in order
for the quantum theory to be single valued with the multivalued classical
action. For noncompact groups there are no topological restrictions for $k$.
The theory possesses the affine symmetry $\hat G\times\hat G$ which entails an
infinite number of conserved currents \cite{Witten3},\cite{Knizhnik}. The
latter can be derived from the basic currents $J$ and $\bar J$,
\begin{eqnarray}
J&\equiv&J^at^a=-{k\over2}g^{-1}\partial g,\nonumber\\ & & \\
\bar J&\equiv&\bar J^at^a=-{k\over2}\bar\partial
gg^{-1},\nonumber\end{eqnarray}
satisfying the equations of motion
\begin{equation}
\bar\partial J=0,~~~~~~~~~\partial\bar J=0.\end{equation}
In eqs. (2.2) $t^a$ are the generators of the Lie algebra ${\cal G}$ associated
with the Lie group $G$,
\begin{equation}
[t^a,t^b]=f^{ab}_ct^c,\end{equation}
with $f^{ab}_c$ the structure constants.

WZNW models based on compact groups are well understood when the level $k$ is
positive integer. For nonnegative integer $k$ a positive definite Hilbert
space, encompassing all the states of the conformal field theory, is defined by
representations of the unitary affine algebra
\cite{Witten3},\cite{Knizhnik},\cite{Gepner}. At the same time, the situation
with negative integer $k$ is far from being well understood. It is clear that
the
theory is no longer unitary because there arise negative normed states in the
spectrum. This is also true for WZNW models on noncompact groups. These
theories are nonunitary due to indefinite Killing metric. However, it has been
noticed by Dixon et al. \cite{Dixon} that the coset $SU(1,1)/U(1)$, which
involves a noncompact group, gives rise to a unitary CFT with a positive
definite Hilbert space \cite{Dixon},\cite{Bars}. Although the Hilbert space of
the affine $SU(1,1)$ algebra may not be positive definite, one may still
construct a positive definite Hilbert space after the projection provided by
the compact $U(1)$ subgroup. It has been argued that the same procedure of
gauging out the maximal compact subgroup of a given noncompact group should
lead
to unitary CFT's in general case \cite{Bars}.

According to Witten's conjecture, WZNW models on compact group manifolds have
to be understood as WZNW models on noncompact group manifolds after appropriate
analytic continuation (Wick rotation) of the compact group to a noncompact
group. The point to be made is that analytic continuation does not spoil
the hermicity condition of the Virasoro generators corresponding to the
affine-Sugawara stress-energy tensor of the WZNW model. Indeed, one can check
that $L^{\dag}_n=L_{-n}$ holds before and after analytic continuation, where
$L_n$ are generators of the Virasoro algebra (see definition of $L_n$ below).
For
generators of the affine Lie algebra this is not true. This observation allows
us to guess that unitary Virasoro representations of the analytically continued
theory have to be unitary representations of the original model before analytic
continuation.

In particular, the noncompact group $SU(1,1)$ can be thought of
as being analytically continued from the compact $SU(2)$. Apparently, an
analytically continued affine algebra will inherit the same level of the
affine algebra of the compact affine Lie group. The WZNW model on the latter
requires the level to be integer. Therefore, the distinctive feature of WZNW
models on noncompact groups obtained by analytic continuation is that their
levels are integer. Because of this fact, our interest in WZNW models on
noncompact groups will be restricted to those having integer level.

Moreover, we will mainly discuss the WZNW model on $SU(1,1)$ at negative level
$k$. This model will be thought of as being obtained by analytic continuation
of the  WZNW model on $SU(2)$ at negative level $k$. Let us discuss the
spectrum of the WZNW model on the noncompact group $SU(1,1)$. The ground states
in our model are the states which are annihilated by the modes $J^a_{n>0}$,
where
\begin{equation}
J^a_n=\oint{dw\over2\pi i}~w^nJ^a(w).\end{equation}
These states will fall into representations of the global $SU(1,1)$ algebra and
since we are requiring unitarity, these will first of all be unitary
representations. All such unitary irreducible representations have been
classified \cite{Bargmann} and since the algebra is noncompact, they will be
infinite dimensional. Denoting the ground state by $|0;j,m\rangle$,
where\footnote{Our convention is
\begin{eqnarray}
f^{21}_3=f^{23}_1=f^{31}_2=1,~~~~g^{ab}={1\over2}f^{bd}_cf^{ca}_d=
\mbox{diag}(-1,-1,1),~~~~g_{ac}g^{cb}=\delta^b_a.\nonumber\end{eqnarray}}
\begin{equation}
g_{ab}J^a_0J^b_0|0;j,m\rangle=j(j+1)|0;j,m\rangle,~~~~~~J^3_0|0;j,m\rangle=m
|0;j,m\rangle,\end{equation}
we have the following nontrivial possibilities
\begin{eqnarray}
&(1)&~~C^0_j~:~j=-{1\over2}~+~i\kappa~~\mbox{and}~~m=0,\pm1,\pm2,...,\nonumber\\
&(2)&~~C^{1/2}_j~:~=-{1\over2}~+~i\kappa~~\mbox{and}~~m=\pm{1\over2},
\pm{3\over2},...,\nonumber\\
&(3)&~~E^0_j~:~-{1\over2}\le j<0~~\mbox{and}~~m=0,\pm1,\pm2,...,\nonumber\\
&(4)&~~D^+_j~:~j=-{1\over2},-1,-{3\over2},...~~\mbox{and}~~m=-j,-j+1,...,
\nonumber\\
&(5)&~~D^-_j~:~j=-{1\over2},-1,-{3\over2},...~~\mbox{and}~~m=j,j-1,...,\nonumber
\end{eqnarray}
where $\kappa$ is real and non-zero. The first three classes,
$C^0_j,~C^{1/2}_j$ and $E^0_j$ are termed continuous, since there the
eigenvalue
of the Casimir operator is a continuous parameter. The last two representations
$D^{\pm}_j$ are called discrete.

We also will be interested in a nonunitary finite dimensional representation of
$SU(1,1)$ with $j=$ half-integer and $-j\le m\le j$. This representation
contains both positive and negative norm states. We will denote this
representation $\Phi^j_m$.

It has been proven \cite{Dixon},\cite{Bars} that the projection of the classes
$C^0_j,~C^{1/2}_j,~E^0_j,~D^{\pm}_j$ onto the $SU(1,1)/U(1)$ coset leads to the
unitary representations of the Virasoro algebra. We are going to exhibit that
there are
more unitary representations of the Virasoro algebra associated with the
$SU(1,1)/U(1)$ coset. These representations will be shown to originate from the
nonunitary finite dimensional representation $\Phi^j_m$ at $j=1$.

For our purposes, it will be convenient to make use of the Lagrangian
formulation of coset constructions \cite{Gawedzki}-\cite{Hwang}. A generic
$G/H$ coset can be described as a combination of ordinary conformal WZNW models
and the ghost-like action,
\begin{equation}
S_{G/H}=S_{WZNW}(g,k)~+~S_{WZNW}(h,-k-2c_V(H))~+~S_{Gh}(b,c,\bar b,\bar c),
\end{equation}
where $h$ takes values on the subgroup $H$ of $G$, $c_V(H)$ is defined
according to
\begin{equation}
f^{il}_kf^{jk}_l=-c_V(H)g^{ij},~~~~i,j,k,l=1,2,...\dim H.\end{equation}
The last term in eq. (2.7) is the contribution from the ghost-like fields,
\begin{equation}
S_{Gh}=\mbox{Tr}\int d^2z~(b\bar\partial c~+~\bar b\partial\bar
c).\end{equation}

The physical states are defined as cohomology classes of the nilpotent
operator $Q$,
\begin{equation}
Q=\oint{dz\over2\pi i}~\left[:c_a(\tilde
J^a+J^a_H):(z)~-~{1\over2}f^{ab}_c:c_ac_bb^c:(z)\right],\end{equation}
where we have used the following notations
\begin{eqnarray}
J_H&=&-{k\over2}g^{-1}\partial g|_H,\nonumber\\ & & \\
\tilde J&=&{(k+2c_V(H))\over2}h^{-1}\partial h.\nonumber\end{eqnarray}
Here the current $J_H$ ia a projection of the ${\cal G}$-valued current $J$ on
the subalgebra ${\cal H}$ of ${\cal G}$.

We have already mentioned that unitary Virasoro representations corresponding
to
the unitary classes of $SU(1,1)$ can be extracted by projecting out the $U(1)$
compact subgroup of $SU(1,1)$. From the point of view of the Lagrangian
approach all
these representations belong to $\mbox{Ker}Q/\mbox{Im}Q$. Let us turn to the
finite dimensional nonunitary representation $\Phi^j_m$. Let us take the
adjoint representation $\Phi^1_m$ which we will denote $\phi^a$, $a=1,2,...,
\dim G$. The ground state corresponding to this operator has the conformal
dimension
\begin{equation}
\Delta_\phi={c_V(G)\over k+c_V(G)}.\end{equation}
The unitarity conditions, which come out from the infinite dimensional
representations, require \cite{Bars}
\begin{equation}
k<-c_V(G).\end{equation}
If condition (2.13) is satisfied, $\Delta_\phi<0$. Hence, a Virasoro
representation on this
highest weight state will be nonunitary. This amounts to the statement that at
level zero the finite dimensional representation $\phi^a$ does not lead to
unitary Virasoro representations.

Let us consider the level one descendant state of the nonunitary representation
of the affine Lie algebra. Namely \cite{Soloviev4},
\begin{equation}
|O^L\rangle=O^L(0)|0\rangle,~~~~O^L(z)=L_{ab}:J^a(z)\phi^b(z):.
\end{equation}
Here the normal ordered product is defined according to \cite{Soloviev1}
\begin{equation}
O^L(z)=L_{ab}\oint{dw\over2\pi i}~{J^a(w)\phi^b(z)\over w-z},\end{equation}
where the product in the numerator of the integrand is understood as an
operator product expansion (OPE). The contour in eq. (2.15) goes anticlockwise
around point $z$. It is easy to see that the given product does not contain
singular terms provided the matrix $L_{ab}$ is symmetrical\footnote{Indeed, the
field $\phi$ is a WZNW primary vector. Therefore, its OPE with the affine
current $J$ is as follows
\begin{eqnarray}
J^a(w)\phi^b(z)={f^{ab}_c\over
w-z}\phi^c(z)~+~reg.~terms.\nonumber\end{eqnarray}
After substitution of this formula into eq. (2.15), one can see that only
regular terms will contribute provided $L_{ab}$ is a symmetrical matrix.}.
According to definition (2.5) of the affine generators, one can present the
state $|O^L\rangle$ as follows
\begin{equation}
|O^L\rangle =L_{ab}J^a_{-1}|\phi^b\rangle.\end{equation}

This state is no longer a highest weight vector of the affine Lie algebra.
At the same time, $|O^L\rangle$ is a highest weight vector of the
Virasoro algebra. Indeed, one can check that
\begin{eqnarray}
L_0|O^L\rangle&=&\Delta_0|O^L\rangle,\nonumber\\ & & \\
L_{m>0}|O^L\rangle&=&0.\nonumber\end{eqnarray}
Here the Virasoro generators $L_n$ are given by the contour integrals
\begin{equation}
L_n=\oint{dw\over2\pi i}~w^{n+1}T(w),\end{equation}
where $T(w)$ is the holomorphic component of the affine-Sugawara stress-energy
tensor of the conformal WZNW model,
\begin{equation}
T(z)={g_{ab}:J^aJ^b:(z)\over k+c_V(G)}.\end{equation}

In eqs. (2.17), $\Delta_O$ is the conformal dimension of the operator $O^L$. It
is not difficult to find that
\begin{equation}
\Delta_O=1~+~{c_V(G)\over k+c_V(G)}.\end{equation}
{}From this formula it is clear that when condition $k\leq-2c_V(G)$ (which is
consistent with (2.13)) is fulfilled the
conformal dimension of $O^L$ is in the range between 0 and 1, i.e. it is
positive. The Virasoro central charge is also positive. Indeed,
\begin{equation}
c_{WZNW}(k)={k\dim G\over k+c_V(G)}=\dim G~-~{c_V(G)\dim G\over k+c_V(G)}>\dim
G>1.\end{equation}
Thus, the operator $O^L$ lies in the unitary range of the Kac-Kazhdan
determinant \cite{Kac} and, hence, it provides a unitary representation of the
Virasoro algebra. However, it is easy to verify that the given operator is not
annihilated by $Q$. Therefore, $O^L$ does not belong to the physical subspace
of the gauged WZNW models. Fortunately, there is a way to modify the operator
$O^L$ without spoiling its properties so that it will belong to
$\mbox{Ker}Q/\mbox{Im}Q$. We will do this modification for the case of the
$SU(1,1)/U(1)$ gauged WZNW model.

First of all, let us restrict ourselves to $O^L$ with a diagonal matrix
$L_{ab}=\lambda_ag_{ab}$ (no summation over indices). Note that for $SU(2)$ the
diagonal form of $L_{ab}$ is the normal one, but in the case of $SU(1,1)$ there
are additional nondiagonalizable forms. In this context, the equation
\begin{equation}
L_{ab}=\lambda_ag_{ab}\end{equation}
is a certain restriction which reminds us that the group $SU(1,1)$ originates
from $SU(2)$ via analytic continuation.

Now let us define a new operator
\begin{equation}
\hat O^L=O^L~+~N~\tilde J^3\phi^3,\end{equation}
where $\tilde J^3$ is the current associated with the compact subgroup
$H=U(1)$,
\begin{equation}
\tilde J^3(w)\tilde J^3(z)={-k/2\over(w-z)^2}~+~reg.~terms.\end{equation}
Here the minus comes from the fact that the level of the affine algebra $\hat
{\cal H}$ acquires the opposite sign to the level of $\hat{\cal G}$. The
constant $N$ is to be determined from the condition
\begin{equation}
Q|\hat O^L\rangle=0,\end{equation}
where the nilpotent charge $Q$ is given by
\begin{equation}
Q=\oint{dw\over2\pi i}~:c~(\tilde J^3~+~J^3):(z).\end{equation}
The second term in eq. (2.10) vanishes in eq. (2.26) because $U(1)$ is Abelian.

It is more convenient to present $Q$ in terms of modes
\begin{equation}
Q={\sum^{+\infty}_{n=-\infty}}:c_{-n}(\tilde J^3_n~+~J^3_n):.
\end{equation}
A canonical choice for the vacuum state of the ghost Fock space is an $SL(2,C)$
invariant state $|0\rangle_{Gh}$, which is annihilated by $L_{Gh,n}$ for
$n=0,\pm1$. This requires that \cite{Karabali_Schnitzer}
\begin{eqnarray}
c_n|0\rangle_{Gh}&=&0,~~~n\ge1,\nonumber\\ & & \\
b_n|0\rangle_{Gh}&=&0,~~~n\ge0.\nonumber\end{eqnarray}
Acting with $Q$ on $|\hat O^L\rangle$ one finds
\begin{equation}
c_0(\tilde J^3_0+J^3_0)|\hat O^L\rangle~+~c_{-1}(\tilde J^3_1+
J^3_1)|\hat O^L\rangle=0.\end{equation}
This condition yields two equations
\begin{eqnarray}
\lambda_1&=&\lambda_2,\nonumber\\ & & \\
N&=&\lambda_3-{2\over k}(\lambda_1+\lambda_2),\nonumber\end{eqnarray}
where $k$ obeys condition (2.13).

Thus, we have found an operator which is annihilated by $Q$. In other words, we
have proved that the operator $\hat O^L$ with $\lambda_i$ and $N$ given by eqs.
(2.30) belongs to the physical subspace $\mbox{Ker}Q/\mbox{Im}Q$. Therefore,
this operator $\hat O^L$ has to be considered on the same footing with the
unitary infinite dimensional representations of $SU(1,1)$. However, we have not
proven that the fusion algebra between $\hat O^L$ and other operators,
commuting
with $Q$, will be free from negative normed operators. In fact, in the large
$|k|$ limit it is not difficult to prove that the $\mbox{Ker}Q/\mbox{Im}Q$
space is evidently unitary \cite{Bars}. More discussion of the unitarity of
$\mbox{Ker}Q/\mbox{Im}Q$ will be given in the next section.

\section{Fusion rules of $\hat O^L$}

It is obvious that the BRST invariant operator $\hat O^L$ and the operator
$O^L$ share one and the same conformal dimension given by eq. (2.20). Of
course, the conformal dimension of $\hat O^L$ is defined with respect to the
Virasoro operator $\hat L_0$ of the gauged WZNW model. The point to be made is
that the unitarity condition in eq. (2.13) indicates that $\hat O^L$ is a
relevant operator. That is,
\begin{equation}
\Delta_{\hat O}=1~+~{c_V(G)\over k+c_V(G)}<1.\end{equation}
Moreover, in the large $|k|$ limit, $\hat O^L$ becomes a relevant quasimarginal
operator. This fact makes the operator $\hat O^L$ of a great interest
because it can be used as a perturbing operator on the given CFT. In the case
under consideration the CFT is the $SU(1,1)/U(1)$ coset which has the target
space interpretation of the two dimensional euclidean black hole\footnote{The
black hole is euclidean because the $U(1)$ subgroup is chosen to be compact.
For the noncompact $U(1)$ the target space geometry of the $SU(1,1)/U(1)$ coset
corresponds to the lorentzian black hole \cite{Witten1},\cite{Dijkgraaf}.}
\cite{Witten1},\cite{Dijkgraaf}. Therefore, perturbations of the $SU(1,1)/U(1)$
coset would amount to perturbations of the euclidean two dimensional black
hole. We will study given perturbations by the operator $\hat O^L$ in the next
section.

Clearly, operators $\hat O^L$ with different arbitrary diagonal matrices
$L_{ab}$ obeying conditions (2.30) will give rise to Virasoro primary vectors
with the same conformal dimension given by eq. (3.31). However, their fusion
algebras may be different. From now on we would like to focus on a particular
subclass of operators $\hat O^L$ which satisfy the following fusion
\begin{equation}
\hat O^L\cdot\hat O^L=[\hat O^L]~+~[I]~+...~,\end{equation}
where the square brackets denote the contributions of $\hat O^L$ and identity
operator $I$ and the corresponding descendants of $\hat O^L$ and $I$, whereas
dots stand for all other admitted operators with different conformal
dimensions. In fact, we will show that only BRST-exact operators and operators
of irrelevant conformal dimensions are admitted to
replace dots in eq. (3.32).

We are going to translate the fusion algebra (3.32)
in an equation for the matrix $L_{ab}$. To this end, it is
convenient to introduce the following operator
\begin{equation}
\Psi^{ab}(z)\equiv J^{(a}_{(-1)}\phi^{b)}(z)=\oint{dy\over2\pi
i}{J^{(a}(y)\phi^{b)}(z)\over y-z},\end{equation}
where indices $a,~b$ are symmetrized. Obviously,
\begin{equation}
O^L={1\over2}L_{ab}\Psi^{ab}.\end{equation}

To start with, let us compute the following OPE
\begin{equation}
\phi^c(w)\Psi^{ab}(z)={C^{ab,c}_d\over(w-z)^{\Delta_\phi}}\phi^d(z)~+~...~,
\end{equation}
where dots stand for all other operators with different conformal dimensions.
To proceed we also have to calculate
\begin{equation}
\phi^a(w)\phi^b(z)={\sum_I}(w-z)^{\Delta_I-2\Delta_\phi}~C^{ab}_I[\Phi^I(z)],
\end{equation}
where $[\Phi^I]$ are conformal classes of all Virasoro primaries $\Phi^I$
arising in the fusion of two $\phi$'s. It is convenient to set $z$ to zero in
eq. (3.36). Then after acting on the $SL(2,C)$ vacuum, eq. (3.36) yields
\begin{equation}
\phi^a(w)|\phi^b\rangle=w^{-\Delta_\phi}~C^{ab}_c|\phi^c\rangle~+~...
\end{equation}
Acting with $J^k_0$ on both sides of eq. (3.37), we find
\begin{equation}
[J^k_0,\phi^a(w)]|\phi^b\rangle~+~f^{kb}_c\phi^a(w)|\phi^c\rangle=C^{ab}_c
f^{kc}_d|\phi^d\rangle~+~...~,\end{equation}
where we took into account the following quantization conditions
\begin{equation}
J^a_0|\phi^b\rangle=f^{ab}_c|\phi^c\rangle,~~~~~~J^a_{m\ge1}|\phi^b\rangle=0.
\end{equation}
The first commutator on the left hand side of eq. (3.38) can be calculated
according to
\begin{eqnarray}
[J^k_n,\phi^a(w)]&=&\oint_w{dz\over2\pi i}~z^nJ^k(z)\phi^a(w)\nonumber\\
&=&\oint_w{dz\over2\pi i}~{z^nf^{ka}_d\phi^d(w)\over z-w}=w^nf^{ka}_d\phi^d(w),
\end{eqnarray}
where one has to use the definition of $J^a_n$ in eq. (2.5).

Finally, we arrive at the following consistency condition
\begin{equation}
f^{ka}_cC^{cb}_d~+~f^{kb}_cC^{ac}_d=C^{ab}_cf^{kc}_d.\end{equation}
Obviously, the solution to this equation is
\begin{equation}
C^{ab}_c=A~f^{ab}_c,\end{equation}
where $A$ is arbitrary constant whose value is to be fixed by appropriate
normalization.

By applying the operators $J_n^a$ to both sides of eq. (3.37) one can fix order
by order all the higher power terms on the right hand side of eq. (3.37).
One more term on the right hand side of eq. (3.37) will be found later on.

Now we turn to the OPE given by eq. (3.35). Again acting with $J^k_0$ on both
sides of eq. (3.35) we obtain
\begin{equation}
[J^k_0,\phi^c(w)]|\Psi^{ab}\rangle~+~\phi^c(w)J^k_0|\Psi^{ab}\rangle
=w^{-\Delta_\phi}
C^{ab,c}_df^{kd}_e|\phi^e\rangle~+~...\end{equation}
The commutator on the left hand side of eq. (3.43) can be computed according to
formula (3.40). Whereas
\begin{equation}
J^k_0|\Psi^{ab}\rangle
=J^k_0J^{(a}_{-1}|\phi^{b)}=f^{k(a}_d|\Psi^{db)}\rangle.\end{equation}

All in all, we find the consistency condition
\begin{equation}
f^{k(a}_dC^{db),c}_n~+~f^{kc}_dC^{ab,d}_n=C^{ab,c}_df^{kd}_n,\end{equation}
which yields the following solution
\begin{equation}
C^{ab,c}_d={A\over2}~(f^{bc}_ef^{ea}_d~+~f^{ac}_ef^{eb}_d),\end{equation}
where one can show that $A$ is the same normalization constant as in eq.
(3.42).

Now let us compute
\begin{equation}
\Psi^{ab}(z)|\Psi^{cd}\rangle=z^{-2\Delta_O+\Delta_\phi}~C^{ab,cd}_n|\phi^n
\rangle~+~
z^{-\Delta_O}~C^{ab,cd}_{mn}|\Psi^{mn}\rangle~+~...\end{equation}
The coefficient $C^{ab,cd}_n$ can be found from the equation
\begin{equation}
[J^k_0,\Psi^{ab}(z)]|\Psi^{cd}\rangle~+~\Psi^{ab}(z)J^k_0|\Psi^{cd}\rangle
=z^{-2\Delta_O+\Delta_\phi}~C^{ab,cd}_nf^{kn}_e|\phi^e\rangle~+~...
\end{equation}
The latter gives rise to the consistency condition
\begin{equation}
f^{k(a}_eC^{eb),cd}_n~-~C^{ab,(ed}_nf^{c)k}_e=C^{ab,cd}_ef^{ke}_n.
\end{equation}
This yields
\begin{equation}
C^{ab,cd}_e={A\over4}\left[(f^{af}_ef^{bd}_nf^{cn}_f+f^{af}_ef^{bc}_nf^{dn}_f+
f^{cf}_ef^{ad}_nf^{bn}_f+f^{df}_ef^{ac}_nf^{bn}_f)~+~(a\leftrightarrow
b)\right].\end{equation}
Further, the coefficient $C^{ab,cd}_{mn}$ is fixed from
\begin{equation}
[J^k_1,\Psi^{ab}(z)]|\Psi^{cd}\rangle~+~\Psi^{ab}(z)J^k_1|\Psi^{cd}\rangle=
z^{-\Delta_O}~C^{ab,cd}_{mn}J^k_1|\Psi^{mn}\rangle~+~...\end{equation}
Here one has to use the following relations
\begin{equation}
J^k_1|\Psi^{ab}\rangle=J^k_1J^{(a}_{-1}|\phi^{b)}\rangle=-f^{k(a}_df^{b)d}_e
|\phi^e\rangle~+~{k\over2}g^{k(a}|\phi^{b)}\rangle.\end{equation}
Whereas the commutator on the left hand side of eq. (3.51) is given by
\begin{equation}
[J^k_1,\Psi^{ab}(z)]=-f^{k(a}_df^{b)b}_e\phi^e(z)~+~{k\over2}g^{k(a}
\phi^{b)}(z)~+~z~f^{k(a}_d\Psi^{db)}(z).\end{equation}

Taking into account eqs. (3.52), (3.53) and (3.35), we come to the following
consistency condition
\begin{eqnarray}
&-&f^{k(a}_ff^{b)f}_eC^{cd,e}_p~+~{k\over2}g^{k(a}C^{cd,b)}_p~-~
f^{k(c}_ff^{d)f}_eC^{ab,e}_p~+~{k\over2}g^{k(c}C^{ab,d)}_p~+~
f^{k(a}_eC^{b)e,cd}_p\nonumber\\
&=&C^{ab,cd}_{mn}({k\over2}g^{k(m}\delta^{n)}_p~-~f^{k(m}_ff^{n)f}_p).
\end{eqnarray}
Bearing in mind the expressions for the coefficients $C^{ab,c}_d,~C^{ab,cd}_n$,
one can compute the coefficient $C^{ab,cd}_{mn}$ from eq. (3.54).

Now it becomes clear that in order to have the operator $O^L$ on the right hand
side of the OPE given by eq. (3.32), we have to impose the following condition
\begin{equation}
L_{ab}L_{cd}C^{ab,cd}_{mn}\sim L_{mn}.\end{equation}
Taking into account eq. (3.54) we obtain
\begin{eqnarray}
&&\left({k\over2}g^{k(m}\delta^{n)}_p~-~f^{k(m}_ff^{n)f}_p\right)L_{mn}\nonumber
\\ & & \\
&=&L_{ab}L_{cd}\left[{k\over2}(g^{k(a}C^{cd,b)}_p+g^{k(c}C^{ab,d)}_p)-
f^{k(a}_ff^{b)f}_eC^{cd,e}_p-f^{k(c}_ff^{d)f}_eC^{ab,e}_p+
f^{k(a}_eC^{b)e,cd}_p\right].\nonumber\end{eqnarray}
This is the equation which yields matrices $L_{ab}$ giving rise to the
following OPE
\begin{equation}
O^LO^L=[O^L]~+~[I]~+~...~,\end{equation}
where dots stand for conformal classes of Virasoro primaries of irrelevant
conformal dimensions. Indeed, it is transparent that the equation (3.56) is
invariant under adjoint transformations of the matrix $L_{ab}$. These
transformations are generated by $J^a_0$. In virtue of this fact, the right
hand side of eq. (3.57) must belong to the equivalence class (or the orbit)
generated by the adjoint action of the global group $G$. Within this class the
OPE in eq. (3.57) is a scalar under global $G$ transformations. There are no
other
relevant Virasoro primary operators but $O^L$ and $I$ which possess the given
property. Hence, the OPE for operators $O^L$ whose matrices $L$ obey the
equation (3.56) is closed on $O^L$ and $I$, and other admitted operators with
irrelevant conformal dimensions (as $:(O^L)^2:$).

The curious fact is that in the limit $k=0$, the obtained equation (3.56) goes
to the master Virasoro equation \cite{Halpern}. However, this limit is beyond
validity of the unitarity condition in eq. (2.13).

We cannot go on with the operator $O^L$ because it does not belong to the
physical subspace of the $SU(1,1)/U(1)$ gauged WZNW model. The BRST invariant
operator is $\hat O^L$ given by eq. (2.22). The additional term in $\hat O^L$
changes the conditions on the matrix $L_{ab}$. In order to take into account
these modifications, we have to compute the following symmetrized OPE
\begin{equation}
\phi^{(a}(z)|\phi^{b)}\rangle=A~C^{ab}_{mn}|\Psi^{mn}\rangle~+~...~,
\end{equation}
where we have already extracted the normalization constant $A$ from the
coefficient. Acting with $J^k_1$ on both sides of eq. (3.58), we obtain
\begin{equation}
[J^k_1,\phi^{(a}(z)]|\phi^{b)}\rangle=A~C^{ab}_{mn}J^k_1|\Psi^{mn}\rangle~+~...
\end{equation}
Now taking into account formula (3.40), we arrive at the following consistency
condition
\begin{equation}
-f^{k(a}_ef^{b)e}_c=C^{ab}_{mn}\left[{k\over2}g^{k(m}\delta^{n)}_c~-~
f^{k(m}_ef^{n)e}_c\right],\end{equation}
from which one can find the coefficient $C^{ab}_{mn}$. In fact, we are
interested only in the coefficient $C^{33}_{mn}$ which is given by
\begin{equation}
C^{33}_{mn}={1\over(k-1)^2-9}\left(\begin{array}{ccc}
k&0&0\\
0&k&0\\
0&0&-1\\
\end{array}\right).\end{equation}
Here
\begin{equation}
k\ne-2,\end{equation}
which is the condition of invertibility of the matrix
$({k\over2}g^{k(m}\delta^{n)}_c-f^{k(m}_ef^{n)e}_c)$ for the case $SU(1,1)$.

Now we can write down the modified equation for the matrix $\hat L_{ab}$
\begin{eqnarray}
&&\left({k\over2}g^{k(m}\delta^{n)}_p~-~f^{k(m}_ff^{n)f}_p\right)~\left(\hat
L_{mn}
+{k\over2}AN^2C^{33}_{mn}\right)\nonumber
\\ & & \\
&=&\hat L_{ab}\hat
L_{cd}\left[{k\over2}(g^{k(a}C^{cd,b)}_p+g^{k(c}C^{ab,d)}_p)-
f^{k(a}_ff^{b)f}_eC^{cd,e}_p-f^{k(c}_ff^{d)f}_eC^{ab,e}_p+
f^{k(a}_eC^{b)e,cd}_p\right],\nonumber\end{eqnarray}
where the constant $N$ is given by eqs. (2.29).

The point to
be made is that in the derivation of the equation (3.63) we did not use the
mode
expansions for conformal operators. These expansions, which in principle can be
defined, are not much of use because we are dealing with operators of rational
conformal dimensions. At the same time, the method of consistency conditions we
have employed here allows us to handle OPE's both for positive and negative
values of $k$.

In what follows our interest will be in the limit $k\to-\infty$. In particular,
we will need to know only the large $|k|$ solution of the equation (3.63).
First
of all note that in this limit, $C^{33}_{mn}\to0$. Therefore, when
$k\to-\infty$ there is no difference between eqs. (3.56) and (3.63). Moreover,
in the large $|k|$ limit the equation (3.63) reduces to the following
\begin{equation}
g^{k(m}\delta^{n)}_p\hat L_{mn}=\hat L_{ab}\hat L_{cd}\left(g^{k(a}C^{cd,b)}_p+
g^{k(c}C^{ab,d)}_p\right)~+~{\cal O}(1/k).\end{equation}
A solution to this equation which fits the conditions in eqs. (2.30) is
\begin{equation}
\hat L_{ab}={g_{ab}\over2A}~+~{\cal O}(1/k).\end{equation}
It is convenient to normalize $\hat L$ to
\begin{equation}
\hat L_{ab}={\sqrt{1\over2}}g_{ab}~+~{\cal O}(1/k).\end{equation}

At the given value of $\hat L$ the operator $\hat O^L$ will satisfy the OPE in
eq. (3.32). Since eq. (3.32) is a gauge invariant version of eq. (3.57), there
will be only gauge invariant extensions of the
operators emerging in eq. (3.57) as well as $Q$-exact operators. Thus,
modulo $Q$-exact terms the
OPE given by eq. (3.32) must be closed on operators $\hat O^L$ and $I$, and
other operators of irrelevant conformal dimensions. In the case of fusion
between $\hat O^L$ and operators from the unitary representations of $SU(1,1)$,
the fusion algebra will be closed on the unitary representations. Indeed, we
first can consider fusion between $O^L$ and the unitary representations.
Because $O^L$ does not change the tensor structure with respect to the global
group $SU(1,1)$, all fusion of $O^L$ have to be closed on global
representations of $SU(1,1)$. In the case of the BRST invariant operator $\hat
O^L$, a part of the fusion algebra can be deduced from the fusion algebra of
$O^L$ by the BRST procedure. In addition to the BRST symmetrized terms there
may appear $Q$-exact terms in the fusion of $\hat O^L$.
All potential operators of irrelevant conformal
dimensions in eq. (3.32) will belong to the unitary range of the Kac-Kazhdan
determinant. Thus, modulo $Q$-exact terms the operator
$\hat O^L$ along with the unitary representations form the closed
unitary fusion algebra. However, the whole space $\mbox{Ker}Q/\mbox{Im}Q$ is
still not unitary. Indeed, the condition of annihilation
by the BRST operator $Q$
is not sufficient to get rid of all nonunitary states. For example, the state
$|\phi^3\rangle$ obeys the BRST symmetry, i.e. $Q|\phi^3\rangle=0$. Thus,
$|\phi^3\rangle$ belongs to $\mbox{Ker}Q/\mbox{Im}Q$. However,
the conformal dimension of $\phi^3$ is nonpositive. Therefore, the descendant
state $L_{-1}|\phi^3\rangle$ is negative normed. This amounts to the
nonunitarity of the ``physical'' subspace of the $SU(1,1)/U(1)$ gauged WZNW
model. Hence, the procedure of gauging out the maximal compact subgroup of the
noncompact group does not automatically lead to the unitarity of the gauged
WZNW model.

\section{Relevant perturbations}

In the previous sections we have exhibited that there are relevant operators in
the space of cohomological classes of the BRST operator of
the gauged $SU(1,1)/U(1)$ model. We have shown that
these operators are highest weight vectors of unitary Virasoro representations
and that their OPE's are closed without introducing new relevant conformal
operators. These properties allow us to make use of the operators $\hat O^L$ to
perform renormalizable perturbations on the $SU(1,1)/U(1)$ coset.

To this end, we first consider relevant
perturbations of the nonunitary $SU(1,1)$ WZNW model. The perturbative theory
is described by the following action
\begin{equation}
S(\epsilon)=S^{*}_{WZNW}~-~\epsilon~\int d^2z~
O^{L,\bar L}(z,\bar z),
\end{equation}
where the first term on the right hand side of eq. (4.67) is the conformal
action of the WZNW model on $SU(1,1)$,
whereas the second term is the perturbation by the operator
\begin{equation}
O^{L,\bar L}(z,\bar z)= O^L(z)~{\bar O}^{\bar L}(\bar z).\end{equation}
Note that within perturbation theory one can apply the theorem of holomorphic
factorization to understand the factorization in eq. (4.68).

{}From now on we will be interested in the limit $k\to-\infty$. This is the
classical limit for the WZNW model on $SU(1,1)$. Correspondingly the operator
$O^{L,\bar L}$ can be presented in the form
\begin{equation}
O^{L,\bar L}(z,\bar z)\buildrel\k\to-\infty\over\longrightarrow
G_{\mu\nu}~\partial x^{\mu}\bar\partial x^{\nu}~+~B_{\mu\nu}~\partial
x^{\mu}\bar\partial x^{\nu},\end{equation}
where $x^\mu$ are coordinates on the group manifold $SU(1,1)$, whereas the
metric and the antisymmetric tensor are given by
\begin{eqnarray}
G_{\mu\nu}&=&-{k\over8}\hat L_{ab}\hat{\bar L}_{\bar a\bar b}~\phi^{b\bar
b}e^a_{(\mu}\bar e^{\bar a}_{\nu)},\nonumber\\ & & \\
B_{\mu\nu}&=&-{k\over8}\hat L_{ab}\hat{\bar L}_{\bar a\bar b}~\phi^{b\bar
b}e^a_{[\mu}\bar e^{\bar a}_{\nu]}.\nonumber\end{eqnarray}
Here $e^a_\mu$ and $\bar e^{\bar a}_\mu$ define left- and right-invariant
Killing vectors respectively. Note that when $\hat L=\hat{\bar L}$,
$B_{\mu\nu}=0$.

Thus, the operator $O^{L,\bar L}$ corresponds to the energy density of the
nonlinear sigma model with metric and antisymmetric field given by eqs. (4.70).
The renormalizability of the gauged sigma model together with eq. (3.56) will
ensure the following OPE
\begin{equation}
O^{L,\bar L}~O^{L,\bar L}=[O^{L,\bar
L}]~+~[I]~+~...~,\end{equation}
where dots stand for operators with irrelevant conformal dimensions. This OPE
agrees with eq. (3.57).

We proceed to calculate the renormalization beta function associated with the
coupling $\epsilon$. Away of criticality, when $\epsilon\ne0$, the
renormalization group equation is given by (see e.g. \cite{Cardy})
\begin{equation}
{d\epsilon\over dt}\equiv\beta=(2-2\Delta_O)\epsilon~-~\pi C\epsilon^2~+~{\cal
O}(\epsilon^3),\end{equation}
where the conformal dimension $\Delta_O$ is given by eq. (2.20), whereas the
coefficient $C$ is to be computed from the three point function
\begin{equation}
\langle O^{L,\bar L}(z_1,\bar z_1) O^{L,\bar L}(z_2,\bar z_2)
O^{L,\bar L}(z_3,\bar
z_3)\rangle=C||O||^2~\Pi^3_{i<j}{1\over|z_{ij}|^{2\Delta_O}}\end{equation}
with $z_{ij}=z_i-z_j,~\bar z_{ij}=\bar z_i-\bar z_j$. Here
\begin{equation}
||O||^2=\langle O^{L,\bar L}(1) O^{L,\bar L}(0)\rangle.\end{equation}

In the large $|k|$ limit one can take into account eq. (3.64) to obtain
\begin{equation}
||O||^2={k^2\over4\dim G}L^{ab}L_{ab}\bar L^{\bar a\bar b}\bar L_{\bar a\bar
b}=k^2\left({3\over16}~+~{\cal O}(1/k)\right),\end{equation}
where the factor $(1/\dim G)$ stems from the normalization of $\phi^{a\bar a}$
\cite{Soloviev2}
\begin{equation}
\langle\phi^{a\bar a}(1)\phi^{b\bar b}(0)\rangle={g^{ab}g^{\bar a\bar
b}\over\dim G}.\end{equation}
Whereas from eq. (3.56) it follows that
\begin{equation}
C=1.\end{equation}

With the given $C$ we can find a nontrivial fixed point of the beta function in
eq. (4.72):
\begin{equation}
\epsilon^{*}=-{2c_V(G)\over\pi k}=-{4\over\pi k}.\end{equation}
At this value of $\epsilon$ the theory in eq. (4.67) becomes a new CFT whose
Virasoro central charge can be estimated by the Cardy-Ludwig formula
\cite{Cardy}
\begin{equation}
c(\epsilon^{*})=c_{WZNW}~-~{(2-2\Delta_O)^3||O||^2\over C^2}~+~
\mbox{higher in $1/k$
terms}.\end{equation}
We find
\begin{equation}
c(\epsilon^{*})={3k\over k+2}~+~{12\over k}~+~{\cal O}(1/k^2).\end{equation}

It turns out that the CFT with the given Virasoro central charge can be
identified with an exact CFT due to the following relation
\begin{equation}
{(-k)3\over(-k)+2}={3k\over k+2}~+~{12\over k}~+~{\cal O}(1/k^2).\end{equation}
Note that $k$ is thought of as being negative integer, so that $-k$ is positive
integer. Thus, the expression on the left hand side of eq. (4.80) coincides
with the Virasoro central charge of the WZNW model on the compact group $SU(2)$
at positive level\footnote{Formally, the formula for the Virasoro central
charge of
the $SU(1,1)$ WZNW model at positive level also fits eq. (4.81). However,
for the given CFT there are no unitary representations in the spectrum due to
the unitarity condition (2.13). At the same time, the perturbation by the
operator $O^{L,\bar L}$ must preserve the unitarity of the positive normed
representations of $SU(1,1)$ in the course of perturbation. Since we know that
such unitary representations exist, we make our choice between the two
options in favour of the WZNW model on the compact group manifold.}.
In other words, the perturbation by the operator
$O^{L,\bar L}(z,\bar z)$ gives rise to the following renormalization group
flow\footnote{The finite shift in the level of the WZNW model on the compact
group can be seen within the fermi-bose equivalence of the non-Abelian Thirring
model \cite{Soloviev4}. Due to this shif the level of the unitary WZNW model
may take value 1, 2 and so on. }
\begin{equation}
SU(1,1)_{-|k|<-4}
\longrightarrow SU(2)_{|k|-4}.\end{equation}

This is a curious result. Indeed, from the point of view of the target
space-time geometry the left hand side of the flow in eq. (4.82)
corresponds to the noncompact group manifold. Whereas the right hand side of
the flow describes the compact group manifold.
Apparently these two
target spaces have different topologies. Thus, the renormalization
group flow at hands provides a certain mechanism of topology change in the
target space. It is different
from the topology change mechanism of Calabi-Yau manifolds \cite{Aspinwall}.
There topology changes under truly marginal deformations. Now we have
exhibited that relevant perturbations also lead to topology change of the
target
space geometry. This may, perhaps, result in a new type of mirror symmetry.

Now let us turn to the case of the BRST invariant perturbation of the
$SU(1,1)/U(1)$ coset. There are two ways to approach this problem. The first
one is to work out the gauge Ward identities for correlation functions dressed
with the quantum gauge fields. This approach mimics the method of studying
CFT's coupled to 2d gravity \cite{Polyakov},\cite{Klebanov},\cite{Soloviev3}.
Gauge dressed correlators are defined as follows
\begin{equation}
\langle\langle\cdot\cdot\cdot\rangle\rangle=\int{\cal D}A{\cal D}\bar A~\langle
\cdot\cdot\cdot~\mbox{e}^{-\int\mbox{Noether~terms}}\rangle,\end{equation}
where the functional integrals over the gauge fields are computed according to
the Faddeev-Popov method.

The second approach is based on the BRST formalism according to which all gauge
dressed correlation functions are written as correlators of BRST invariant
operators. In particular for $O^{L,\bar L}$ one will have
\begin{equation}
\langle\langle O^{L,\bar L}(z_1,\bar z_1)O^{L,\bar L}(z_2,\bar
z_2)\cdot\cdot\cdot O^{L,\bar L}(z_n,\bar z_n)\rangle\rangle\sim
\langle\hat O^{L,\bar L}(z_1,\bar z_1)\hat O^{L,\bar L}(z_2,\bar
z_2)\cdot\cdot\cdot\hat O^{L,\bar L}(z_n,\bar z_n)\rangle,\end{equation}
where the operator $O^{L,\bar L}$ corresponds to the following gauge invariant
expression
\begin{equation}
O^{L,\bar L}=-{k^2\over4}\left\{\mbox{Tr}(\partial g\bar\partial g^{-1})~+~
2\mbox{Tr}[Ag^{-1}\partial g-\bar A\bar\partial gg^{-1}+\bar Ag^{-1}Ag-A\bar
A]\right\},\end{equation}
with $A=A_z\cdot t^3,~\bar A=A_{\bar z}\cdot t^3$ and $L=\bar L=1$.
Whereas the operator $\hat O^{L,\bar L}$ is defined as follows
\begin{equation}
\hat O^{L,\bar L}(z,\bar z) =\hat O^L(z)~\hat{\bar O^{\bar L}}(\bar
z),\end{equation}
with $\hat O^L$ given by eq. (2.23).

Correspondingly, one can study the renormalization group flows in the presence
of the quantum gauge fields following the two ways just described above. The
important point to be made is that Polyakov has proved that there is no
renormalization of the gauge coupling constant \cite{Polyakov}. Therefore, all
renormalization will amount to the renormalization of the perturbation
parameter $\epsilon$ in the perturbative action
\begin{equation}
S_{gauged}(\epsilon)=S^{*}_{coset}~-~\epsilon~\int d^2z~\hat O^{L,\bar
L}(z,\bar
z),\end{equation}
where $S^{*}_{coset}$ is the
conformal action of the gauged $SU(1,1)/U(1)$ model.

Due to eq. (3.63), the coefficient $C$ in the three point function of the BRST
invariant operator $\hat O$ is equal to one. Therefore, to leading orders in
$\epsilon$, the renormalization group equation for the coupling $\epsilon$
coincides with eq. (4.72). Hence, all fixed points of the perturbed WZNW model
remain critical points of the perturbed gauged WZNW model. Thus, the point
$\epsilon^{*}$ given by eq. (4.78) has to be a conformal point of the theory in
eq. (4.87).  However, correlation functions undergo the gauge dressing which
affects the Virasoro central charge at the infra-red conformal point.  Indeed,
the dressed two-point function is given by
\begin{equation}
\langle\langle O^L(1)O^L(0)\rangle\rangle=\langle\hat O^L(1)\hat
O^L(0)\rangle=k^2\left({1\over8}~+~{\cal O}(1/k)\right).\end{equation}
This expression differs from the one in eq. (4.75). Correspondingly the
Cardy-Ludwig formula changes. The alterations obscure the interpretation of the
CFT at the IR critical point. For example, the formula for the Virasoro central
charge now is given by
\begin{equation}
c(\epsilon^{*})={3k\over k+2}~-~1~+~{8\over k}~+~{\cal O}(1/k^2).\end{equation}
It is not clear to which exact CFT this perturbative central charge may
correspond.

Instead of the $SU(1,1)/U(1)$ coset, one can start with the slightly modified
CFT described by the direct sum of the $SU(1,1)/U(1)$ coset and a free compact
scalar field. The action is
\begin{equation}
S_{{SU(1,1)\over U(1)}\times U(1)}=S_{SU(1,1)/U(1)}~+~{1\over4\pi}\int
d^2z~\partial X\bar\partial X.\end{equation}
Now the space $\mbox{Ker}Q/\mbox{Im}Q$ contains a new relevant operator
\begin{equation}
O_X=\partial X~\phi^3.\end{equation}
Note that the scalar field $X$ is inert under the gauge transformations. There
exists a combination of the operators $\hat O^L$ and $O_X$ which obeys the
fusion algebra in eq. (3.32):
\begin{equation}
O^L_X=\hat O^L~+~E~O_X,\end{equation}
where $E$ is arbitrary constant whose physical meaning will be clarified
presently.  The matrix $L$ depends on $E$ through the consistency condition
which can be derived in a similar way to how  we derived the equation (3.63).

If we choose
\begin{equation}
E=\pm i\sqrt{|k|/2},\end{equation}
then the norm of $O^L_X$ will coincide with eq. (4.75). Therefore, after
perturbation of the CFT in eq. (4.90) by the operator $O^L_X$, the system
arrives at the IR critical point with the following Virasoro central charge
\begin{equation}
c(\epsilon^{*})={2k\over k+2}~+~{12\over k}~+~{\cal O}(1/k^2).\end{equation}
The last formula allows us to identify the given CFT with the $(SU(2)\times
U(1))/U(1)$ coset construction.
All in all, we come to conclusion that at the critical point $\epsilon^{*}$ the
perturbed $(SU(1,1)/U(1))\times U(1)$ coset  coincides with the action of the
gauged $(SU(2)\times U(1))/U(1)$ model
at positive level. This will amount to the following renormalization group flow
\begin{equation}
{SU(1,1)_{k<-4}\over U(1)}\times U(1)\longrightarrow{SU(2)_{|k|-4}\times
U(1)\over
U(1)}.\end{equation}
{}From the point of view of the target space geometry, the left hand side of
eq.
(4.95) describes a two-dimensional electrically charged eucledian black hole
\cite{Ishibashi}, whereas the right hand side of the
flow is the product of the two-dimensional sphere and a circle.  One might
expect to have the last geometry for the equilibrium of two extreme black
holes.  In this connection, the constant $E$ defines electric charge of the
black hole, whereas the equation (4.93) coincides with the condition for
extreme black holes. However, this conjecture has to be more carefully
investigated. The aforementioned effect once again illustrates target
space topology change triggered by relevant perturbations.

\section{Conclusion}

We have started with the analysis of the spectrum of the $SU(1,1)/U(1)$ coset
at negative level and proceeded to define relevant operators corresponding to
highest weight vectors of unitary Virasoro representations. We found that these
representations come into being as level one descendants of the highest weight
vector of the nonunitary (finite dimensional) representation of the noncompact
affine Lie algebra at negative level. We have established that these relevant
operators can be arranged to form the closed fusion algebra.

We have performed the large $|k|$ renormalizable perturbation of the
$(SU(1,1)/U(1))\times U(1)$ coset and found that the perturbed model has a
nontrivial
conformal point. It has been displayed that this perturbative conformal point
corresponds to the $(SU(2)\times U(1))/U(1)$ coset at positive level. Thus, we
have
exhibited the new mechanism of topology change in the target space along the
renormalization group flow. Therefore, it might be interesting to understand
whether or not this topology change is related to a new kind of duality
symmetry
or mirror symmetry of string theory.

There is another quite interesting issue which we left for further
investigation. Namely, one can consider the $N=2$ supersymmetric generalization
of the $SU(1,1)/U(1)$ coset. This theory describes the unitary $N=2$
superconformal discrete series \cite{Lykken}. Our conjecture is that the
perturbation we discussed in the present paper can provide flows between $c>3$
and $c<3$
$N=2$ series \cite{Soloviev}.

\par \noindent
{\em Acknowledgement}: It is a pleasure to thank C. Hull, T. Ortin and S.
Thomas for useful discussions.
I would also like to thank the PPARC and the European Commission Human Capital
and Mobility Programme for financial support.

\end{document}